\documentclass[preprint,12pt,authoryear]{elsarticle}

\usepackage{amssymb}
\usepackage[normalem]{ulem}
\usepackage{color}
\usepackage{siunitx}

\usepackage{lineno}

\journal{XXXXXXX}

\begin{document}

\begin{frontmatter}



\title{Detection of spark discharges in an agitated Mars dust simulant isolated from foreign surfaces}


\author[1]{Joshua M\'endez Harper}
\author[1]{Josef Dufek}
\author[3]{George D McDonald}

\address[1]{Department of Earth Science, University of Oregon, Eugene OR, 97403, USA}
\address[3]{Department of Earth and Planetary Sciences, Rutgers, The State University of New Jersey, Piscataway NJ, 08854, USA}

\begin{abstract}

Numerous laboratory experiments, starting in the Viking Lander era, have reported that frictional interactions between Martian analog dust grains can catalyze electrostatic processes (i.e. triboelectrification). Such findings have been cited to suggest that Martian dust devils and dust storms may sustain lightning storms, glow discharges, and other complex electrostatic phenomena. However, in many cases (if not most), these experiments allowed Martian dust simulant grains to contact foreign surfaces (for instance, the wall of an environmental chamber or other chemically-dissimilar particles). A number of authors have noted that such interactions could produce charging that is not representative of processes occurring near the surface of Mars. More recently, experiments that have identified and corrected for collisions between dust simulants and chemically dissimilar laboratory materials have either failed to replicate near-surface Martian conditions or directly measure discharging. In this work, we experimentally characterize the triboelectrification of a Martian dust simulant resulting from both isolated particle-particle collisions and particle-wall collisions under a simulated Martian environment. We report the direct detection of spark discharges in a flow composed solely of natural basalt grains and isolated from artificial surfaces. The charge densities acquired by the fluidized grains are found to be of order 10\textsuperscript{-6} Cm\textsuperscript{-2} (in excess of the theoretical maximum charge density for the near-surface Martian environment). Additionally, we demonstrate that the interaction of simulant particles with experimental walls can modulate the polarity of spark discharges. Our work supports the idea that small-scale spark discharges may indeed be present in Martian granular flows and may be qualitatively similar to small-scale discharges in terrestrial volcanic vents.

\end{abstract}

\begin{keyword}
 Tribocharging \sep Martian electricity \sep Mars dust storms



\end{keyword}

\end{frontmatter}



\section{Introduction}
Triboelectric charging---that is, electrification through frictional and collisional interactions---is extremely common in granular flows, both natural \citep{aplin2012laboratory, gu2013role, mendez_harper_effects_2016, mendez_harper_electrification_2017, mendez2018tribo} and artificial \citep{lowell1986triboelectrification, lacks_effect_2007, apodaca2010contact, lacks_contact_2011}. The effects of tribocharging can be subtle, like the aggregation of fine volcanic ash high in Earth's atmosphere \citep{james_experimental_2002,telling_experimental_2012, telling2013ash}, or explosively dramatic like the lightning storms that accompany volcanic eruptions \citep{thomas_electrical_2007, bennett_monitoring_2010, arason_charge_2011, behnke_observations_2013, nicora2013actividad, aizawa_physical_2016, cimarelli_multiparametric_2016, behnke_investigating_2018, mendez2018inferring, hargie2018globally}. Given their ubiquity on Earth, triboelectic processes likely operate in other planetary environments with large dust and sand reservoirs, such as Mars, Titan, and, perhaps, a number of exoplanets  \citep{krauss_experimental_2003, aplin2006atmospheric, forward_particle-size_2009, aplin2012laboratory, helling2013ionization, mendez_harper_electrification_2017, mendez2018tribo}. 

\begin{figure}[!]
	\centering
	\includegraphics[width=\columnwidth]{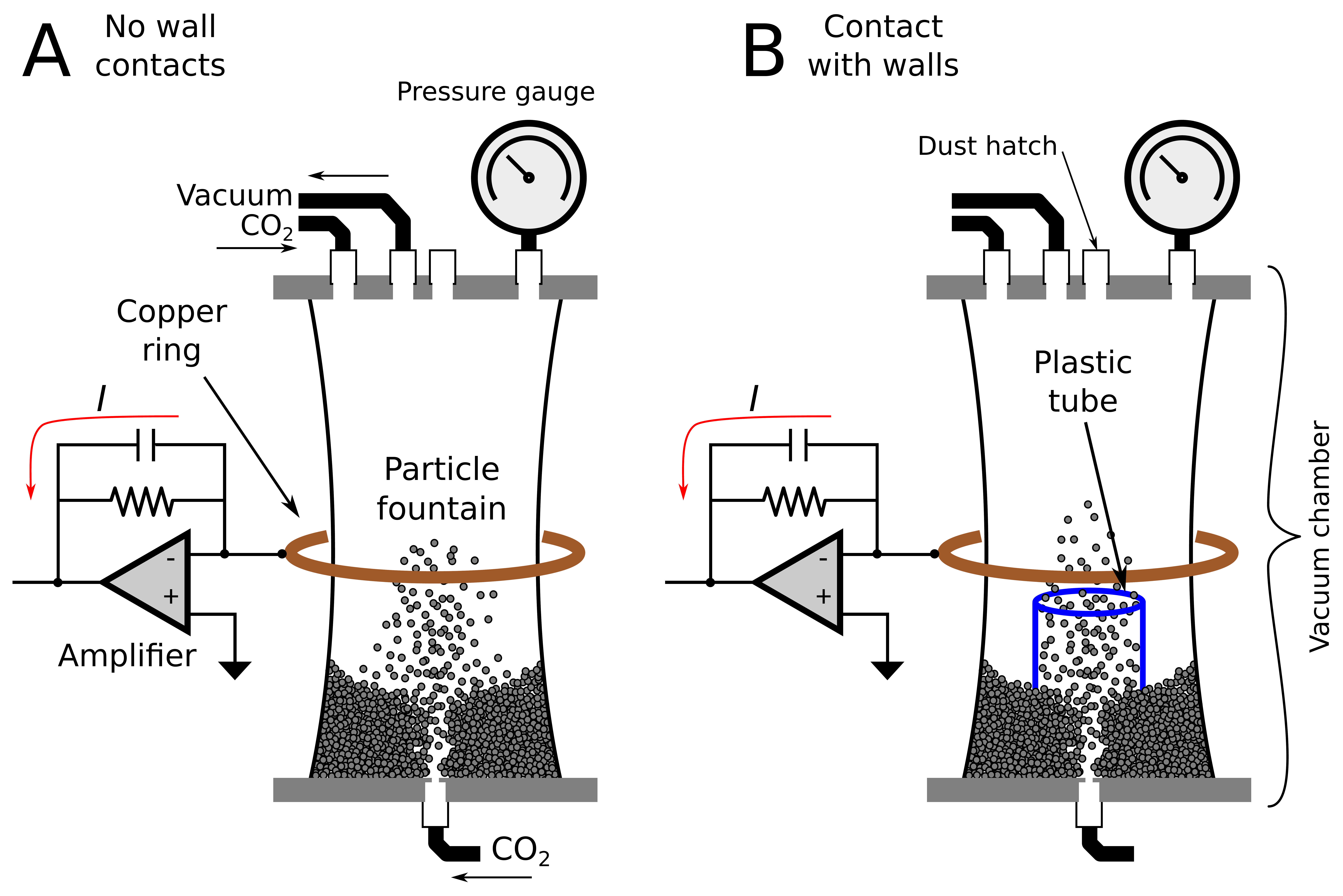}
	\caption{\label{fig1} The setup used in this work to assess frictional electrification of a Martian dust simulant. The system consists of a glass environmental chamber capable of approximating near-surface Martian conditions. Particles in the chamber were charged by fluidizing a bed of basaltic material for periods of 20 minutes. Any discharges in the flow were detected by a electrostatic ring. We also used this setup to characterize the charging resulting from the interaction of particles with a plastic wall. A) Simulant grains undergo only particle-particle interactions. B) Simulant grains interact with an acrylic cylinder as well as with each other.}
\end{figure}  

Theoretical, numerical, and experimental analyses suggest that the low-pressure (4.0 to 8.7 mbar), CO\textsubscript{2} atmosphere at the Martian surface breaks down under smaller electrical stresses than the near-surface terrestrial atmosphere \citep{kok_wind_blown_2009, helling2013ionization,  harrison2016applications,wurm2019challenge,riousset2020scaling}. Indeed, spark discharges could occur at electric fields on the order of a few tens of kV m\textsuperscript{-1} on Mars. Such fields are much weaker than those required to produce breakdown in Earth's near-surface atmosphere of 3 MV m\textsuperscript{-1}. These theoretical Martian breakdown fields are also comparable to those that have been measured in terrestrial dust storms (~100 kV m\textsuperscript{-1}; \cite{stow1969dust, crozier1964electric, farrell2004electric}). The interest in characterizing tribocharging of Martian materials stems from the hypothesis that, like on Earth, triboelectric charging on Mars may have the capacity to modulate a wide range of phenomena. Such processes could include enhanced dust lofting \citep{farrell2004electric,zhai2006, kok_wind_blown_2009} the aggregation of charged grains \citep{krinsley1981,merrison2004electrical}, the photochemical destruction and production of species with implications for the presence of organics and methane \citep{atreya2006b,farrell2006b,tennakone2016contact}, and electrical discharges \citep{melnik1998electrostatic, krauss_experimental_2003, krauss2006modeling, wurm2019challenge}.

 \begin{figure}[!]
	\centering
	\includegraphics[width=5 in]{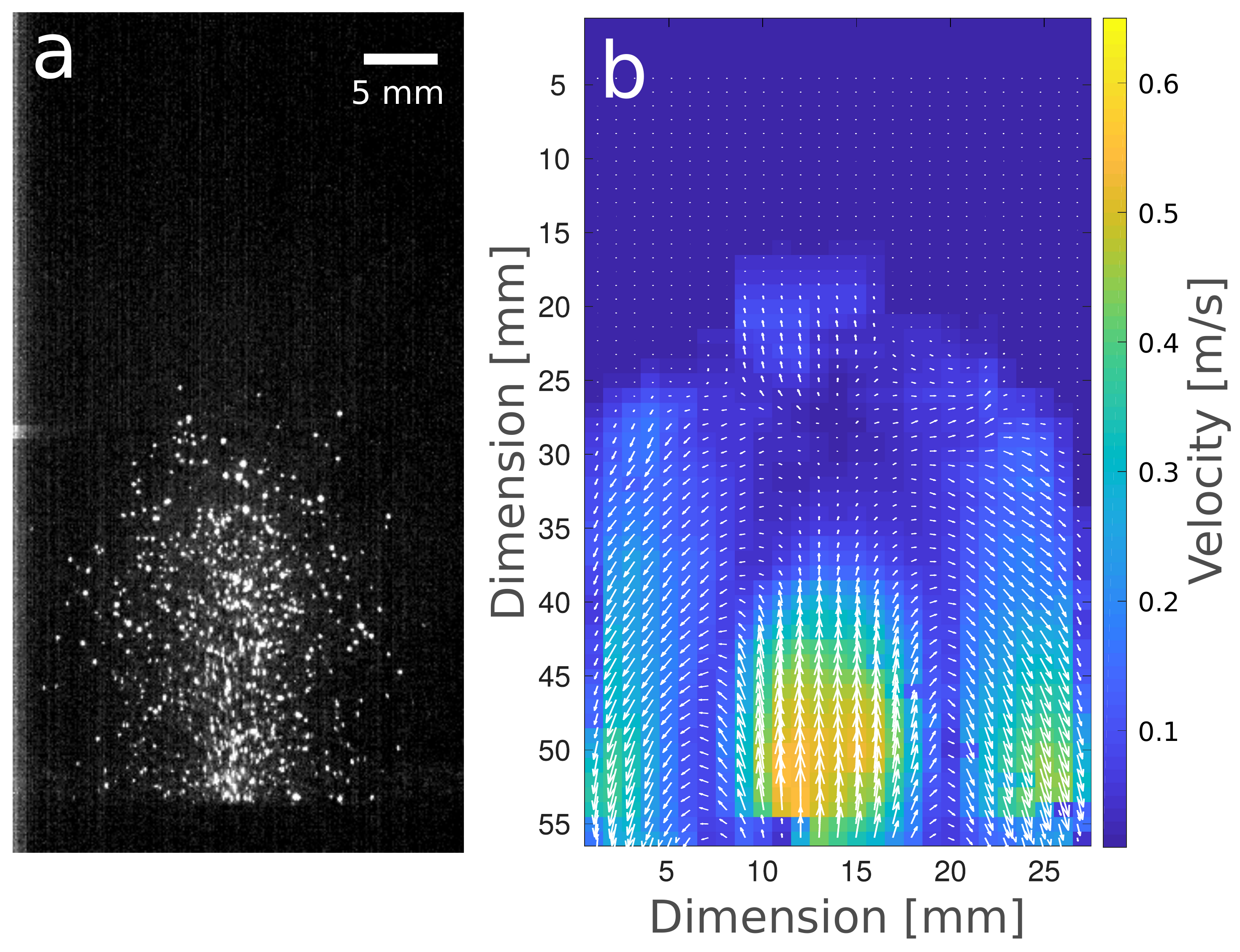}
	\caption{\label{fig2} Fountain dynamics. a) High speed photograph showing fountain illuminated by a sheet laser. b) Velocity distribution in the fountain as measured through particle image velocimetry over 50 frames. }
\end{figure}

In the context of the Martian environment, putative near-surface triboelectricty has been hypothetically discussed since the observation of silts, clays, and aeolian bedforms by the Viking landers \citep{greeley1979}. The subsequent discovery of dust devils by the Viking orbiters \citep{thomas1985} has led to speculation that, as on Earth, dust devils and large scale storms generate electrically charged grains \citep{stow1969dust, crozier1964electric, farrell2004electric}. Indirect evidence for triboelectric charging on the Martian surface has continued to be reported, as well as revisited, over the years. For instance, the Wheel Abrasion Experiment (WAE) as part of the Mars Pathfinder mission's Soujourner rover reported photographic evidence for the adhesion of dust to the rover's wheels in Mars Year (MY) 23. The dust was observed to have adhered preferentially to the aluminum and platinum components of the WAE based on reflectance measurements. An electrostatic origin for this adhesion was suggested but could not be verified since Sojourner did not have the capability to measure on-board electrical charge \citep{ferguson1999mars}. 

Despite tangential evidence, no dedicated electrostatic measurements have been performed \textit{in-situ} on the Martian surface for verification of this phenomenon. However, a diverse set of experiments conducted across the last 40 years have successfully produced both glow and spark discharges in laboratory-scale flows of Martian dust simulants. \cite{eden1973electrical} investigated the triboelectric charging of silicates under an approximated Martian atmosphere by agitating sand in a glass flask. These investigators found that the motion of grains produced both small spark and glow discharges observable in a dark room. Similar experiments were conducted by \cite{mills1977dust}. More recently, \cite{krauss_experimental_2003} performed experiments using the JSC-Mars-1 simulant (weathered basaltic volcanic ash from Mauna Kea) in a polycarbonate chamber. Glow discharges were recorded when the simulant was mobilized by a motorized stirrer. \textcolor{black}{These authors also characterized the triboelectrification of a mixture of basaltic material and glass microspheres falling through a hopper \citep{krauss2006modeling}}. \cite{mazumder2004mars} electrified JSC-Mars-1 simulant in a plastic vial and measured the charge on individual grains using laser Doppler velocimetry.

\begin{figure}[!]
	\centering
	\includegraphics[width=4 in]{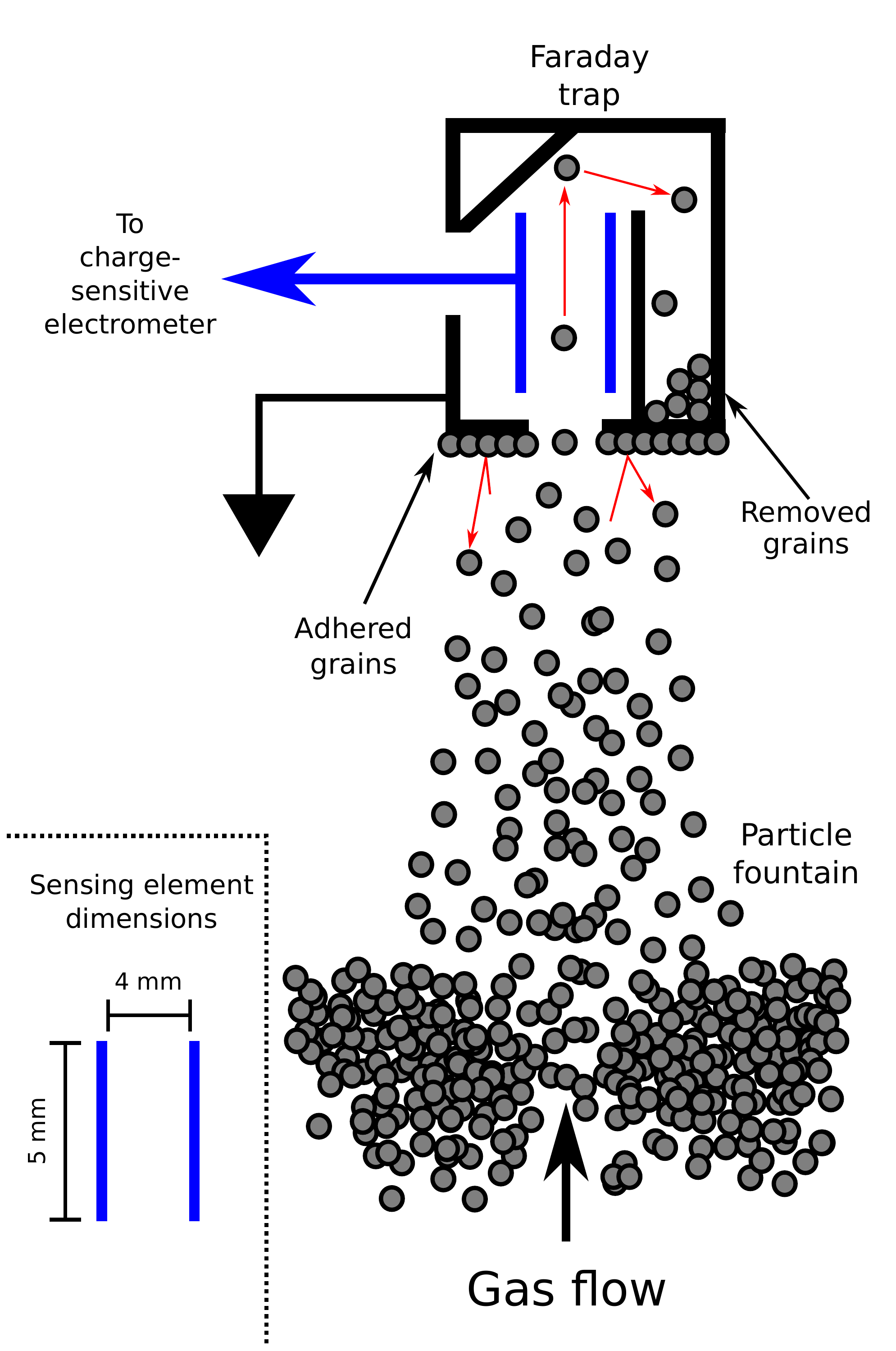}
	\caption{\label{fig3} Schematic of the process used to estimate the charge density on particles in the flow. A Faraday particle trap (FPT) was suspended over the fountain in the vacuum chamber. After concluding the fluidization period (20 min), the inlet pressure was increased, launching particles into the narrow aperture in the bottom of the FPT. The internal sensing element (marked in blue) was small enough to detect the passage of individual grains with diameters down to 100 microns. Once traversed the sensing volume, particles are deflected into a discard bin by an inclined plane. This was done to prevent measured particles (which may have contacted the internal surfaces of the FPT) from becoming re-entrained into the fountain. The outside of the FPT was coated with grains of the same composition to minimize dissimilar material charging between the fountain and the FPT's copper walls}.
\end{figure}  

The insights provided by the laboratory studies summarized above are invaluable. Nevertheless, the direct applicability of the results to \textit{in-situ} electrification on Mars is hindered by the fact that analog materials were allowed to contact surfaces that would not be expected on the Martian surface. The experiments of \cite{eden1973electrical}, \cite{krauss_experimental_2003}, and \cite{mazumder2004mars} involved copious particle-wall interactions. \textcolor{black}{\cite{krauss2006modeling} reduced wall effects, but still used mixtures of natural volcanic material and chemically-dissimilar artificial glass microspheres.} Collisions with container walls, the vanes of a stirrer, or between particles of different compositions can contribute significantly to the overall electrification through two principal mechanisms.

Firstly, the interaction of two chemically-dissimilar materials can produce more intense charging than the interaction between two chemically-identical materials \citep{diaz2004semi, zou2019quantifying, zou2020quantifying}. Such an effect has been described empirically and compiled into a so-called \textit{triboelectric series} by countless authors (e.g. \cite{shaw1917experiments}). Thus, even if single component particle-particle collisions are more frequent than particle-wall collisions in an experiment, the particle-wall interactions may drive the overall electrification of the granular media \citep{saleh2011relevant, xing2020effect}.

Secondly, the segregation and accumulation of charge can be much more efficient during the interaction of surfaces with large discrepancies in size (consider micron-sized particles rubbing against the comparatively large inner surface of a flask) when compared to those produced by binary particle-particle collisions. Dozens of experiments have shown that when two surfaces repeatedly interact, the larger one acquires a positive charge while the smaller one becomes negatively charged \citep{forward_particle-size_2009, waitukaitis2014size,  bilici_particle_2014, lee2015direct}. \textcolor{black}{Typically, the two surfaces are particles of different sizes \citep{lowell1980contact, lowell1986triboelectrification, lacks_effect_2007, lacks_contact_2011}.  However, the same effect would be expected during the repeated interaction between a spherical particle and a large flat plane or the curved interior of a tube, or between two planes rubbing against each other in an asymmetric fashion  (like a knife on a whetstone). What matters here is that there exists a difference in scale between the two contacting surfaces (see \textbf{Section 5.1} and \textbf{Figure 6} in \cite{lacks_contact_2011}).} Thus, in the experiments described above, even if the container and the grains had been the \textit{exact} same chemical composition, the large container walls would have gained a positive charge relative to the small agitated particles. Such size-dependent triboelectric charging may have set up coherent electric fields within experimental apparatuses, readily permitting the low-pressure CO\textsubscript{2} to reach breakdown conditions.

Together, the chemical and geometrical disparities between interacting grains and chamber boundaries could have produced the previously reported triboelectric effects in agitated Martian dust simulants. Indeed, \cite{aplin2012laboratory} have demonstrated that the results presented in \cite{krauss_experimental_2003} and \cite{krauss2006modeling} can be explained entirely by interactions between natural silicate particles and apparatus walls. We note that understanding how Martian materials charge when interacting with human-made materials \textit{does} provide useful insights given the numerous upcoming missions to Mars. However, because the electrification observed in many previous studies arose from experimental artifacts, the results of those works cannot be directly applied to understanding charging in natural Martian systems. \cite{aplin2012laboratory} conclude their paper by encouraging the community to develop experiments that isolate Martian simulants from foreign objects to better understand endemic tribocharging. 

The effects of walls were minimized, if not eliminated, in the charging experiments of JSC-Mars-1 by \cite{forward_particle-size_2009}. These authors used a spouted bed to assess grain charging resulting from particle-particle collisions only. However, those experiments did not approximate the near-surface Martian environment (\cite{forward_particle-size_2009} used a nitrogen environment at 90 mbar), nor did the authors explore the viability of electrical discharges in the spouted bed. More recently, \cite{wurm2019challenge} performed careful experiments in a vibrated bed to study the breakdown conditions for the near-surface Martian environment. To minimize the effects of dissimilar material charging, the authors coated the walls of their setup with particles of the same composition and size. \cite{wurm2019challenge} found that particles acquired charge densities of at least 7-11 $\mu$C m\textsuperscript{-2}. However, those authors used monodisperse samples with relatively large particles (1-2 mm) which provide limited insight into the charging dynamics of material mobilized by aeolian action. 

Here, we investigate triboelectric processes under simulated Martian environments with an experimental setup that eliminates boundary condition effects and allows for verification that particles are not only charging, but also actively discharging during fluidization. We conduct experiments in an 8 mbar CO\textsubscript{2} atmosphere using a Martian dust simulant similar to the Mojave Mars Simulant \citep{peters2008mojave}. We employ a set of sensors that can both detect spark discharges in the flow and directly measure the charge density on individual grains. Additionally, we conduct experiments in which particles are allowed to contact a foreign surface to highlight the differences in charging behavior as compared to that of an isolated flow (which best mimics granular flows in natural settings). Our  measurements suggest that the near surface Martian environment may support small-scale spark discharges resulting from the electrification due solely to particle-particle collisions. Our work also demonstrates that the configuration of the experimental setup is of critical importance when attempting to assess triboelectric processes in planetary environments.

\section{Methods} \label{secMethods}

\subsection{Martian atmosphere simulator and fluidized bed apparatus}

Electrification of a Martian dust simulant through frictional interactions was achieved using a fluidized bed apparatus comparable to that  described in \cite{forward_particle-size_2009} and \cite{mendez_harper_effects_2016}. The device consists of a vertical glass tube (diameter $\sim$ 10 cm) capped off at each end by an aluminum plate. This arrangement forms a pressure/vacuum chamber in which the environment can be finely tuned. \textbf{Figures \ref{fig1}a} and \textbf{b} show the system schematically. Approximately 100 g of a Martian dust simulant (described further on) was inserted into the chamber through a hatch in the top plate. The sample was then sprayed with a \textcolor{black}{bipolar} ExAir charge neutralizing air gun and allowed to sit for 24 hours at ambient conditions to permit any initial charge to dissipate.  After this period of respite, the chamber was evacuated to 1-3 mbar and then purged by refilling the chamber with CO\textsubscript{2} gas to a pressure of 1 bar. Subsequently, the pressure in the chamber was brought down to 8 mbar, close to the average near-surface Martian pressure (4.0 to 8.7 mbar). A closed-loop control program running on an Arduino development board maintained a stable pressure in the chamber. All experiments were conducted at 25\textsuperscript{o} C.

\begin{figure}[!]
	\centering
	\includegraphics[width=\columnwidth]{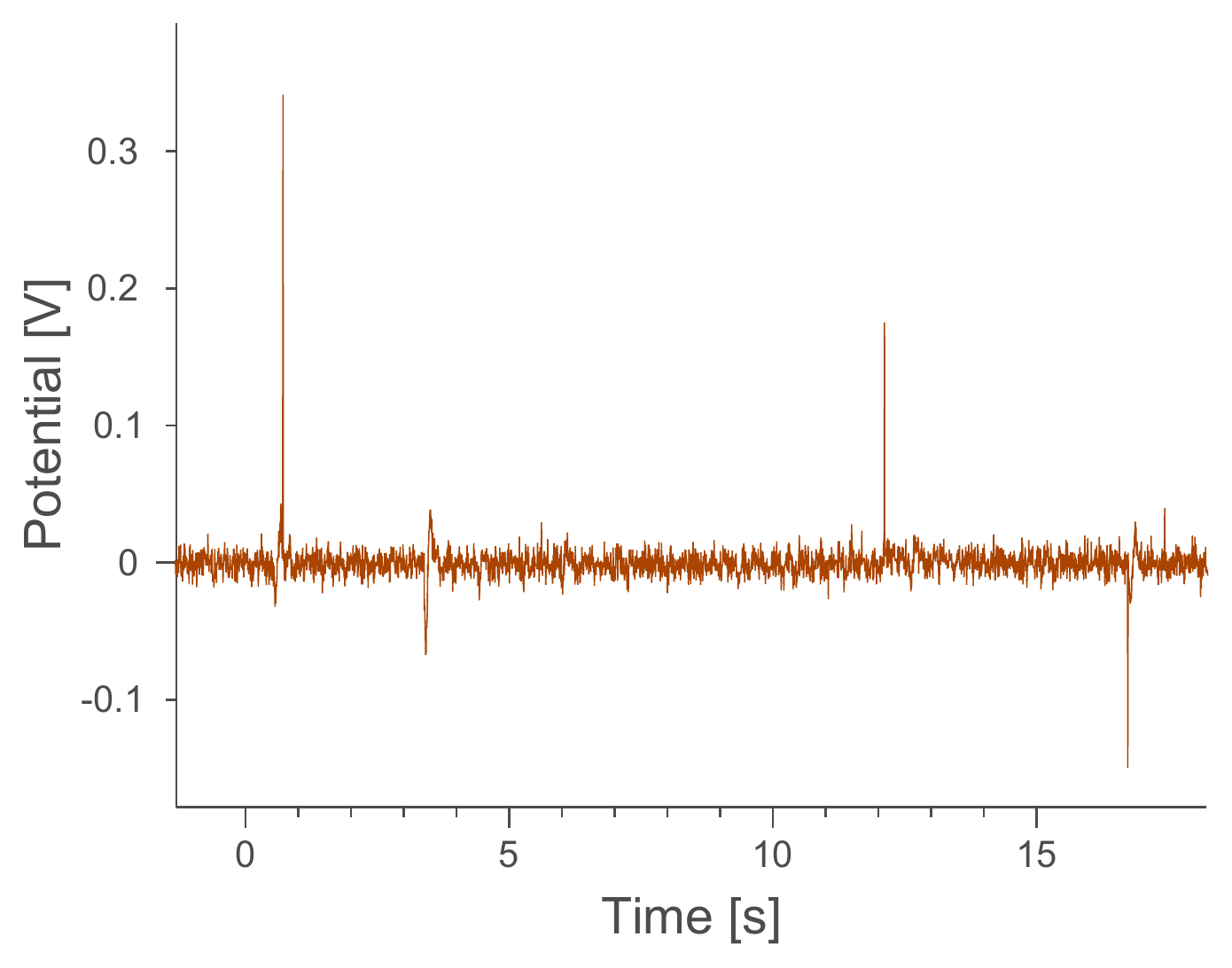}
	\caption{\label{fig4} Typical output waveform from the transimpedance amplifier connected to the electrostatic ring during an experiment involving particle-particle collisions only (the setup in \textbf{Figure \ref{fig1}a}). Sharp transients are discharges in the flow.}
\end{figure}

To fluidize the bed of particles, a jet of CO\textsubscript{2} was forced through a 500 micron hole milled in the bottom end cap. As particles are mobilized, they collide with each other and exchange charge triboelectrically. By filming the fountain at high speed and processing the frames with a particle image velocimetry code, we estimate that the particles had maximum velocities of 0.6 m/s  (see \textbf{Figure \ref{fig2}}). \textcolor{black}{Previous efforts have explored the charging of Martian dust simulants under higher 2 -- 7 m/s laboratory ``wind speeds" in attempts to match the freestream winds characterized by the Viking 1 Lander (measured at a height of 1.6 m above the ground) \citep{krauss_experimental_2003}. Rather than match the freestream velocity, we designed our experiments to approximate the threshold friction velocities for grains at the surface of 0.12 -- 0.43 m/s (using the \textit{in-situ} roughness length of 2.5 $\times 10^{-3}$ m, \cite{hebrard_roughness_2012}). The frictional velocity is the minimum shear velocity required for initiation of motion of grains (i.e. the start of saltation). Thus, our experiment allows us to characterize the triboelectric charging of a Martian simulant at the lowest particle kinetic energies expected on Mars.}

The particle fountain in our experiments had an elevation of ~4 cm above the bed surface (see \textbf{Figure \ref{fig2}}) The inner diameter of the glass tube was large enough that particles did not contact the walls during fountaining. Thus, unlike most previously published experiments, triboelectric charging in this setup stems overwhelmingly from the interaction of chemically-identical surfaces (see \textbf{Figure \ref{fig1}a}). However, to demonstrate the effect of experimental setup on the triboelectrifcation of the simulant, we also fluidized the sample in the presence of a foreign object. For these tests, we placed a hollow acrylic cylinder (diameter 3.8 cm, length 2 cm) concentrically around the jet aperture. Here, particles interacted with each other and a plastic surface which may have produced spurious electrification (\textbf{Figure \ref{fig1}b}).

\begin{figure}[!]
	\centering
	\includegraphics[width=4 in]{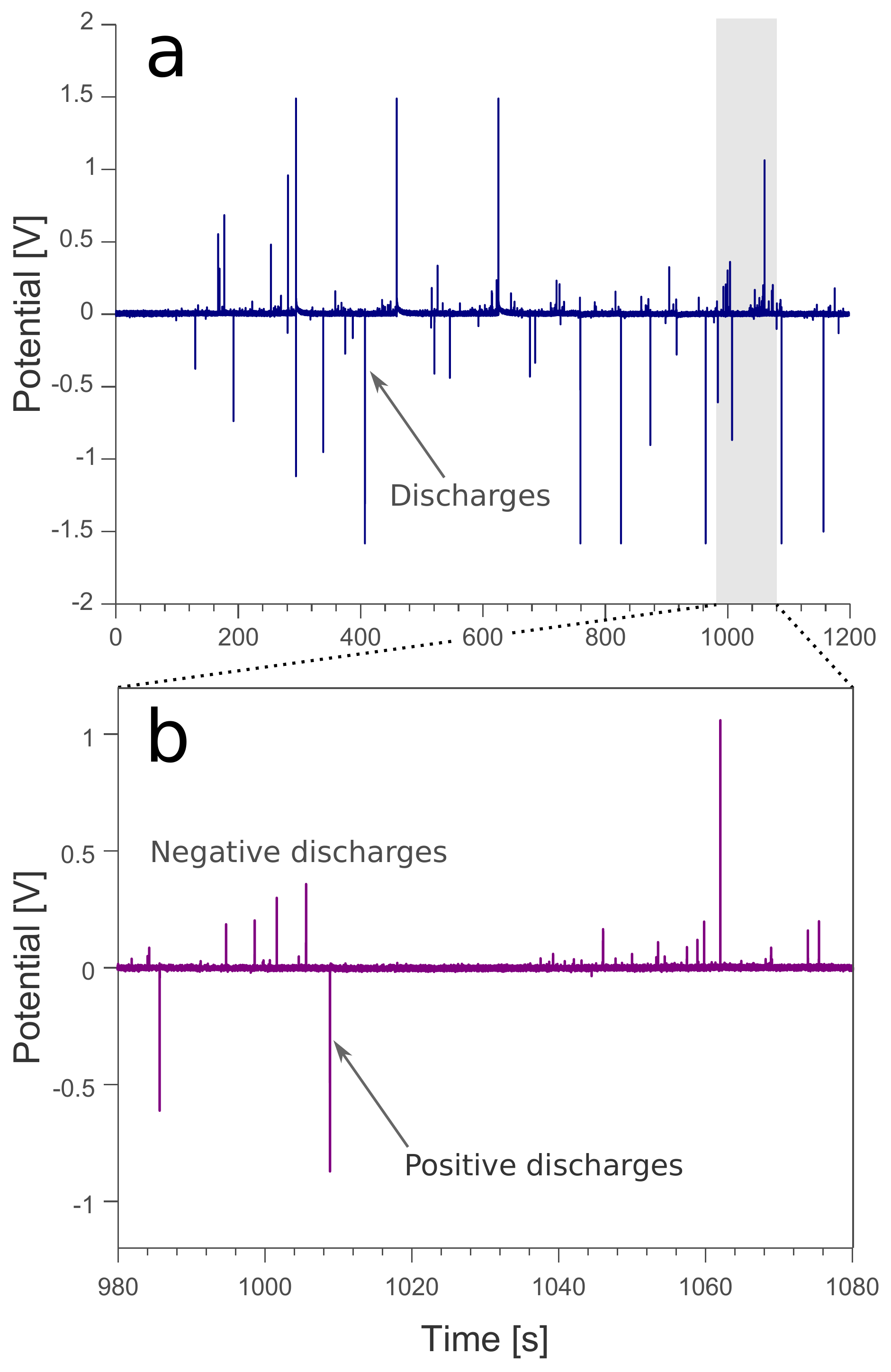}
	\caption{\label{fig5} a) Time series of discharges as measured by the ER in an isolated fountain composed of our Xictli Mars dust simulant. Note the generally bipolar discharge behavior. b) Expanded section from a) to show that negative discharges are more frequent, but neutralize smaller amounts of charge than the large, less frequent positive discharges.}
\end{figure}  

\subsection{Measurement electronics}

We employed two measurement devices to characterize the electrification in the fluidized bed: 1) An \textcolor{black}{electrostatic ring (ER)} capable of detecting spark discharges and 2) a miniature Faraday particle trap (FPT) to measure the charge on individual particles sampled from the flow. We note that the instruments were used in separate experiments. 

The \textcolor{black}{ER consists of a copper ring antenna running along the outside of the glass chamber. When a moving charged particle passes through the cross-sectional area of the ring it produces a time-varying potential in the ring that is related to the velocity of the charge, its magnitude, and position. A precision transimpedance amplifier converts the induced signal in the ring into a measurable voltage. Similar designs have been used to electrostatically monitor the pneumatic transport of granular materials in a non contact fashion \citep{gajewski1999non, gajewski2008electrostatic}}. The output of the ER circuit is monitored for a period 20 minutes after the start of fluidization. We used the ER to characterize the discharge behavior in both the isolated fountain and the fountain with wall contacts (i.e. the experiments with the plastic insert). The ER is rendered schematically in \textbf{Figure \ref{fig1}}.

We employ the FPT to measure the charge density (with units of C m\textsuperscript{-2}) on individual particles in the fluidized particle bed. For these measurements, we replaced the ER and suspended a miniaturized inverted Faraday cup from the top end cap of the pressure chamber (See \textbf{Figure \ref{fig3}}). Here, we only characterized the charge density on particles in an isolated fountain. The granular sample was fluidized for twenty minutes at a constant inlet pressure. After this period, the inlet pressure was marginally increased for a period of 5 minutes, flinging particles near the dilute top of the fountain into the 5 mm-wide aperture of the FPT. This measurement process is capable of measuring the charge on individual particles and is similar to that reported in \cite{watanabe2006measurement, mendez_harper_effects_2016} and \cite{mendez_harper_electrification_2017}. In contrast to the setup involving the ER, some particle-wall collisions do occur with the FPT setup. However we are able to minimize the effects of dissimilar charging while performing these measurements, by invoking two strategies: 1) we coated the exterior surface of the FPT with particles of the same size so that rebounding particles do not contact the copper body of the FPT; and 2) the internal geometry of the FPT is designed to trap particles once charge on them has been measured. Specifically, the top interior wall of the FPT is inclined so that over 80\% of the particles that enter the sensing element are removed, and do not reflect back into the fountain (as quantified through high speed video). The precautions are depicted schematically in \textbf{Figure \ref{fig3}}. The sensing element of the FPT is connected to a charge sensitive preamplifier (built around an LMC6082 operational amplifier) which outputs a voltage proportional to the charge passing through the sensing element with a resolution of 1 pC/V. The fundamental principles of our FPT are based on the design described extensively in \cite{watanabe2006measurement}.

\subsection{Martian dust simulant} \label{secDust}

As a Martian dust simulant, we use crushed olivine basalt from the $\sim$300 AD eruption of the Xictli (meaning ``navel" in Nahuatlahtolli and hispanicized as \textit{Xitle}) volcano (Chichinauhtzin Volcanic Field, Distrito Federal, M\'exico) \citep{delgado1998xitle}. The sample corresponds to Flow 5 and was sourced from an outcrop near parking lot 4 of the Cultural Center of the National Autonomous University of Mexico (19\si{\degree}18'44.7''N, 99\si{\degree}11'02.2''W). In terms of composition, Xictli basalt is similar to the Mojave Mars Simulant (\cite{peters2008mojave}; see \textbf{Table \ref{tb1}}).

\begin{table}   
	\caption{Chemical composition comparison between Mars dust simulants. Concentrations are in wt\%. The  values in parentheses denote the differences in composition of our Xictli sample from the most commonly used simulants. See \cite{allen1998jsc, delgado1998xitle} and \cite{peters2008mojave} for more details.} 
	\begin{center} 
		\begin{tabular}{cccccc} 
			Oxide & JSC-Mars-1 & MM1 & MGS-1 & Xictli, Flow 5\\ 
			\hline 
			SiO\textsubscript{2} & 43.5 \textsubscript{(-7.3)} & 47.9 \textsubscript{(-2.9)} & 41.4 \textsubscript{(-9.4)} &   50.8  \\
			Al\textsubscript{2}O\textsubscript{3} & 23.3 \textsubscript{(7.4)} & 16.7 \textsubscript{(0.8)} & 11.2 \textsubscript{(-4.7)} &   15.9  \\
			TiO\textsubscript{2} & 3.62 \textsubscript{(1.82)} & 1.09 \textsubscript{(-0.71)} &  0.2 \textsubscript{(-1.6)} & 1.8   \\
			FeO\textsubscript{2} \& Fe\textsubscript{2}O\textsubscript{3} & 15.3 \textsubscript{(6.0)} & 10.6 \textsubscript{(1.3)} & 13.3 \textsubscript{(4.0)} &   9.3   \\
			MnO & 0.26 \textsubscript{(0.1)} & 0.17 \textsubscript{(0.01)} & 0.1 \textsubscript{(-0.06)} & 0.16 \\ 
			CaO & 6.2 \textsubscript{(-2.7)} & 10.45 \textsubscript{(1.55)} & 2.2 \textsubscript{(-6.7)} & 8.9 \\
			MgO & 4.2 \textsubscript{(-4.5)} & 6.08 \textsubscript{(-2.62)} & 14.8 \textsubscript{(6.1)} & 8.7 \\
			K\textsubscript{2}O & 0.7 \textsubscript{(-0.32)} & 0.48 \textsubscript{(-0.54)} & 2.3 \textsubscript{(1.28}) & 1.02 \\
			Na\textsubscript{2}O & 2.34 \textsubscript{(-0.99)} & 3.28 \textsubscript{(-0.05)} & 4.3 \textsubscript{(0.97)} & 3.33  \\
			P\textsubscript{2}O\textsubscript{5} & 0.78 \textsubscript{(0.3)} & 0.17 \textsubscript{(-0.31)} & 0.3 \textsubscript{(-0.18)} & 0.48 \\
		\end{tabular} 
	\end{center} 
	\label{tb1}
\end{table}

\begin{table} 
	\caption{Particle size distribution of the crushed Xictli ash used in the experiments with the electrostatic ring based on the particle size distribution of JSC-Mars-1 \citep{allen1998jsc}. Boxed distributions were used in experiments with the FPT.}
	\begin{center} 
		\begin{tabular}{cccccc} 
			Grain size ($\mu$m) &  Wt\%\\ 
			\hline 
			500-1000  & 21.4  \\
			\framebox[1.1\width]{250-500} &  29.5\\
			\framebox[1.1\width]{150-250}  & 20.8  \\
			90-150 &  12.9\\
			45-90  & 9.2  \\
			20-45 &  5.4\\
			$<$20 &  1.3\\
		\end{tabular} 
	\end{center} 
	\label{tb2}
\end{table}

For the experiments using the ER, the Xictli ash was sieved, washed, and dried to obtain a size distribution mimicking that of the JSC-Mars-1 simulant commonly used in other experiments (see \cite{allen1998jsc} and \textbf{Table \ref{tb2}}). For the FPT experiments, we used two samples: one with particles with nominal diameters between 150-250 $\mu$m and another with particles in the range of 250-500 $\mu$m. We did not employ the full polydisperse simulant described in \textbf{Table \ref{tb1}} because 1) we found we could not determine whether or not grains smaller than $\sim$100 $\mu$m  were being effectively trapped by the sensor; 2) the sheer number of small particles traversing the sensor prevented us from assessing the charge on individual grains and; 3) narrow size distributions reduced the uncertainty involved in estimating charge density. 


\section{Results and Discussion}

\subsection{Discharges in the flow and the effect of experimental setup} \label{secResults:subsecdischarges}

\textcolor{black}{Our electrostatic ring detected impulsive  transients suggestive of spark discharges. As noted by \cite{gajewski2008electrostatic}, discharges in a granular flow produce rapid changes in the local electric field, leading to transient potential changes in the ring electrode. Such events were present in all experiments---i.e. these signals} occurred in both the experiments involving isolated particle fountains and those with particle-wall contacts. \textcolor{black}{We meticulously investigated our experimental setup to ensure that such signals were not the result of pump noise or other electromagnetic interference in the lab. We found that the transients only occurred during fountaining in a low pressure atmosphere and were qualitatively similar to those described by \cite{krauss_experimental_2003, krauss2006modeling} and \cite{farrell_is_2015}.} Four discharge signals (at the output of amplifying circuit) \textcolor{black}{from an experiment with an isolated fountain} are exemplified in \textbf{Figure \ref{fig4}}. Discharge events began within a few minutes of the start of fluidization. The magnitude of the signals increased with time and eventually saturated the amplifier circuit (the voltage pulses were clipped at the amplifier's voltage rails).

\subsubsection{Experiment without wall effects}

\textcolor{black}{The presence of discharges in the experiments with the isolated fountain suggests that mobilized sub-millimeter  silicate particles can effectively charge up to the gas breakdown limit under a rarefied CO\textsubscript{2} atmosphere. Because the flow was isolated from other surfaces, charge exchange (i.e. accumulation of charge on grain surfaces) must have occurred between constituent particles only. Consequently, the discharges we detected likely occurred within the fountain's volume where we expect the electric fields to be strongest. Although possible, we believe that discharges between the fountain and the chamber's wall are unlikely given that the wall does not actively charge during an experiment and is always a few centimeters away from particle fountain. In other words, discharges are much more likely to occur between two oppositely-charged grains in close proximity than between a charged grain and a nominally neutral boundary with a greater distance between the two.} 

\begin{figure}[!]
	\centering
	\includegraphics[width=\columnwidth]{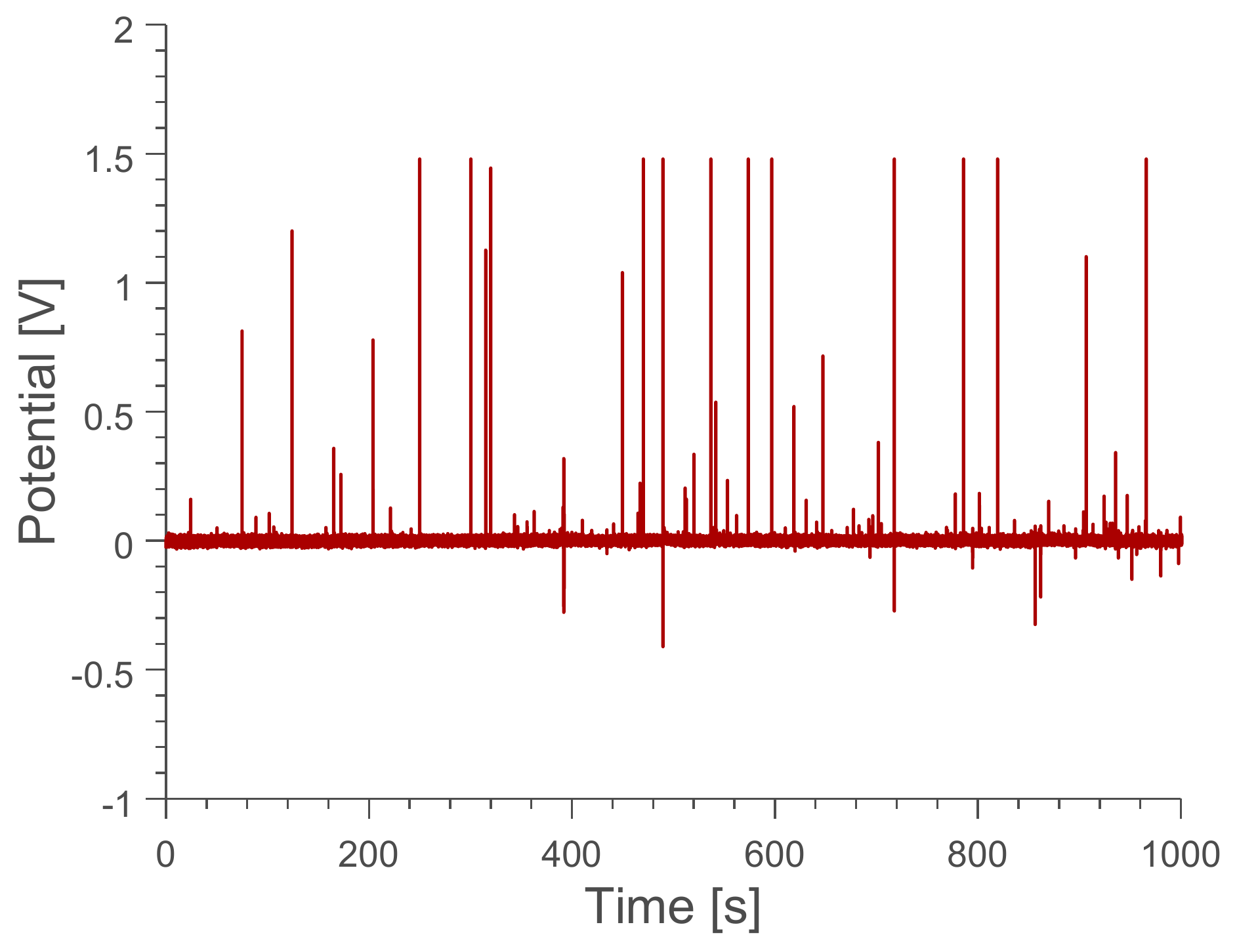}
	\caption{\label{fig6} Typical output waveform from the ER circuit in the experiment involving collisions of the particles with an acrylic cylinder wall (the setup in \textbf{Figure \ref{fig1}b}). Note the overwhelmingly monopolar discharge behavior.}
\end{figure} 


A 20 minute recording of the electrostatic ring's output for a typical isolated fountain experiment is rendered in \textbf{Figure \ref{fig5}a}. Note the overall bipolarity of the discharges.  Neutralization of negative charge produces a positive change in potential, whereas neutralization of positive charge leads to a negative potential change. Interestingly, however, negative discharges (that is, discharges resulting in positive potential changes) are more frequent than positive discharges, but have overall smaller magnitudes (see \textbf{Figure \ref{fig5}b}, which shows an expanded section of \textbf{Figure \ref{fig5}a}). This behavior may have to do with how charge carriers (potentially electrons) are partitioned among particles in a polydisperse flow. As mentioned previously, numerous authors have reported that negative charge becomes preferentially concentrated on the smaller particles in a granular material (including JSC-Mars-1; see \cite{forward_particle-size_2009}). Larger particles are left with positive charges. Because of their relatively inextensive surface areas, smaller particles are able to sustain comparatively smaller amounts of net charge than large particles before the simulated Martian atmosphere breaks down. We posit that the asymmetry in magnitudes between positive and negative discharges are underpinned by these geometrical considerations.  

\textcolor{black}{Over all, discharges occurred at maximum rates of 0.1-1 Hz, lower than those reported by \cite{krauss_experimental_2003} (up tp 15 Hz) and \cite{krauss2006modeling} (up to 7 discharges in one second-long experiments). Furthermore, unlike \cite{eden1973electrical} or \cite{krauss_experimental_2003}, we were unable to detect a glow discharge under darkroom conditions with the unaided eye. We discuss potential reasons for these differences in \textbf{Section \ref{secResults:subsecRate}}.}



\subsubsection{Experiment with wall effects}

\begin{figure}[!]
	\centering
	\includegraphics[width=\columnwidth]{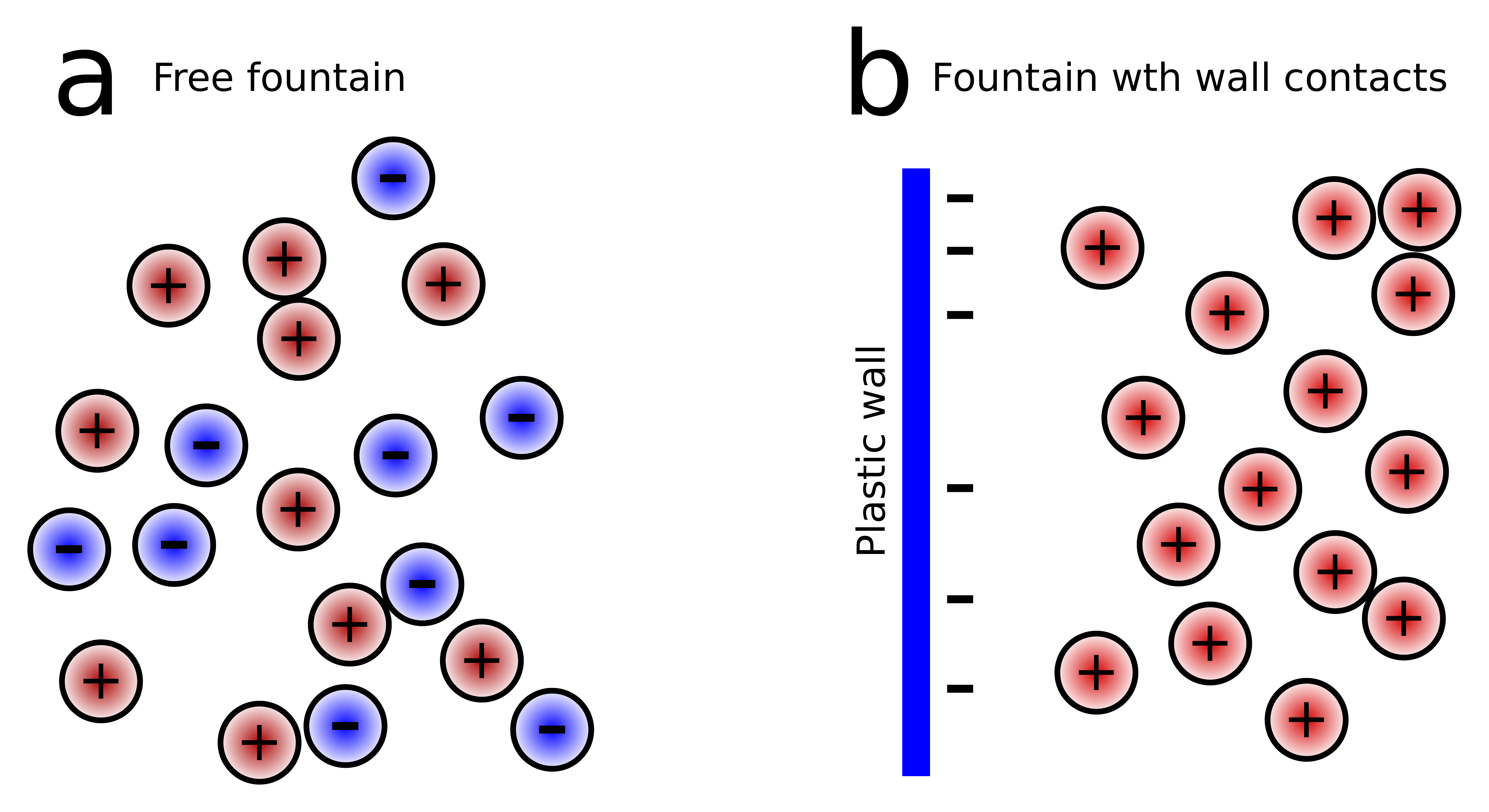}
	\caption{\label{fig7} Conceptual representation of charge partitioning in a) an isolated particle fountain and b) a flow of particles in which grains are allowed to contact a foreign wall. When the flow is isolated, particles exchange charge only with each other (i.e. some particles become positive while others concentrate negative charge). Conversely, when the flow is allowed to interact with a plastic wall, the direction of charge flow is dominated by relative differences in the materials' work functions. In the present case, silicates tend to charge positively whereas plastics charge negatively. This difference in charge accumulation likely underpins the differences in discharge behavior we observe in our spouted bed experiments (See Figures \ref{fig5} and \ref{fig6}).}
\end{figure}

As mentioned in the Methods section, we conducted a second set of spouted bed experiment in which we allowed our simulant to contact a foreign object, specifically an acrylic cylinder. The voltage trace rendered in \textbf{Figure \ref{fig6}} exemplifies the discharge behavior in these experiments. In contrast to the bipolar discharging observed in the free-fountaining experiments, discharges in the experiments with the plastic insert were overwhelmingly negative (i.e. they produced positive voltage changes). This finding is unsurprising given the documented charging behaviors when two or more dissimilar materials are brought into contact (see \cite{matsusaka_triboelectric_2010} and \cite{lacks_contact_2011} for comprehensive reviews). When the fountain is isolated, charge exchange occurs exclusively between constituent grains. Thus, while the charge on any individual grain may be very large, either positive or negative, the net charge of the fountain is zero. Conversely, when the flow is exposed to a boundary of different composition, there is a preferred direction of charge flow determined by the surface work functions of the ash and plastic. In other words, negative charge tends to collect on one of the surfaces, leaving the other with a net positive charge. The direction of charge flow (i.e. which substance will be positive or negative) during frictional or contact interactions has been summarized empirically in the \textit{triboelectric series} \citep{shaw1917experiments}. Silicates tend to be at the top of the series, whereas hydrocarbon-based materials dominate the lower portion. Thus, in the experiments involving the plastic insert, ash particles likely charge positively, whereas the acrylic cylinder accumulates negative charge. The difference in charge accumulation between experiments with and without the plastic insert is presented conceptually in \textbf{Figure \ref{fig7}b} and \textbf{a}. The effective segregation of charge in experiments with the  plastic insert likely accounts for the monopolar discharge behavior shown in \textbf{Figure \ref{fig6}}. Interestingly, the discharge rate did not change significantly as compared to the isolated fountain experiments (in contrast with the polarity). This finding may suggest that ash-plastic contacts here involve similar rates of charge exchange as that in particle-particle contacts. 

To close this section, we would like to highlight two main findings: 1) As far as we are aware, this is the first work to report spark discharges in an agitated bed of Mars dust simulant particles isolated from both wall contacts and interactions with particles of dissimilar compositions. Our work, thus, lends credence to the idea that small scale discharges are plausible in the near-surface Martian dust and sand systems; 2) As in \cite{aplin2012laboratory}, our experiments demonstrate that boundary effects can have profound impacts on the charging and discharge behaviors of Mars dust simulants. Specifically, our experiments show that a bipolar discharge behavior in an isolated fountain readily switches to an overwhelmingly monopolar one when a passive acrylic wall is introduced in the system. Together, our findings stress that future investigations of Mars dust charging should take appropriate precautions to minimize contacts with foreign objects.
 
 \begin{figure}[!]
	\centering
	\includegraphics[width=5 in]{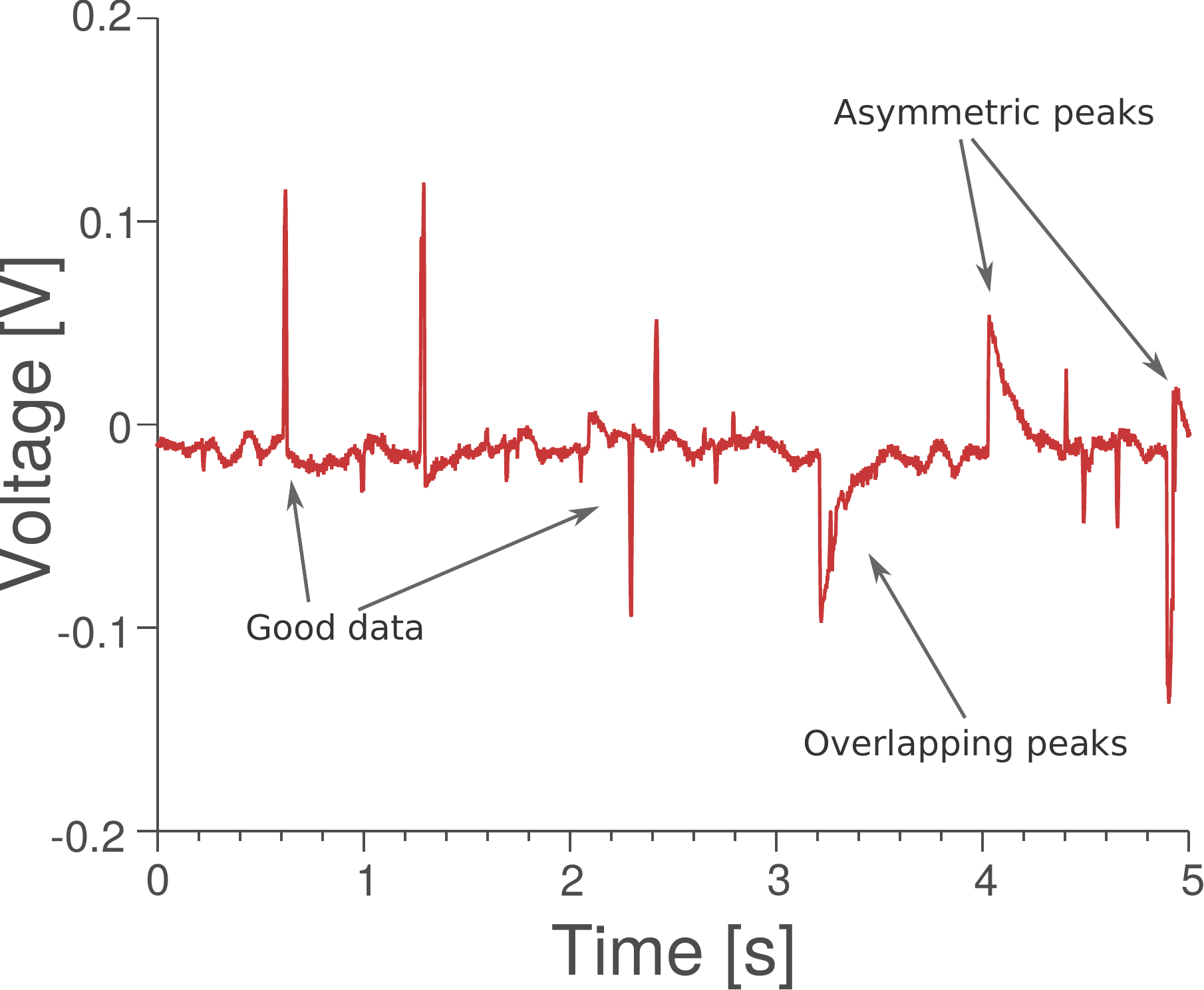}
	\caption{\label{fig8} Voltage trace exemplifying the the output of the charge amplifier circuit connected to the FPT. Each pulse represents the passage of a single particle through the sensing volume of the circuit. Symmetric pulses are representative of particles that traversed the sensing volume without impacting the interior wall of the FPT. Overlapping pulses are representative of more than one particle passing through the sensing volume at a time. We discard asymmetric pulses and overlapping pulses from our dataset.}
\end{figure}  

\subsection{Charge density on particles}

The second set of experiments involved the Faraday particle trap (FPT), allowing us to directly estimate the charge density on individual grains in the flow. As mentioned in the methods section, we only assessed the charge on flows composed of particles with diameters of 150-250 $\mu$m and 250-500 $\mu$m. The time series rendered in \textbf{Figure \ref{fig8}} shows the output voltage of the charge amplifier. Each peak is representative of a single particle entering the FPT. Asymmetric peaks indicate situations in which particles may have contacted the interior walls of the sensing element (thus, the charge entering the FPT is different than the exiting charge). Overlapping peaks result from the passage of more than a single particle through the sensor at a time. We exclude these peaks (indicated in \textbf{Figure \ref{fig8}}) from further analysis. Overall, we analyzed the charge on $\sim$100 particles for each distribution. We note that while the FPT can measure charges as small as 1 fC, microphonic noise from our vacuum pump prevented us from reliably detecting charges smaller than 0.01 pC. 

\begin{figure}[!]
	\centering
	\includegraphics[width=4 in]{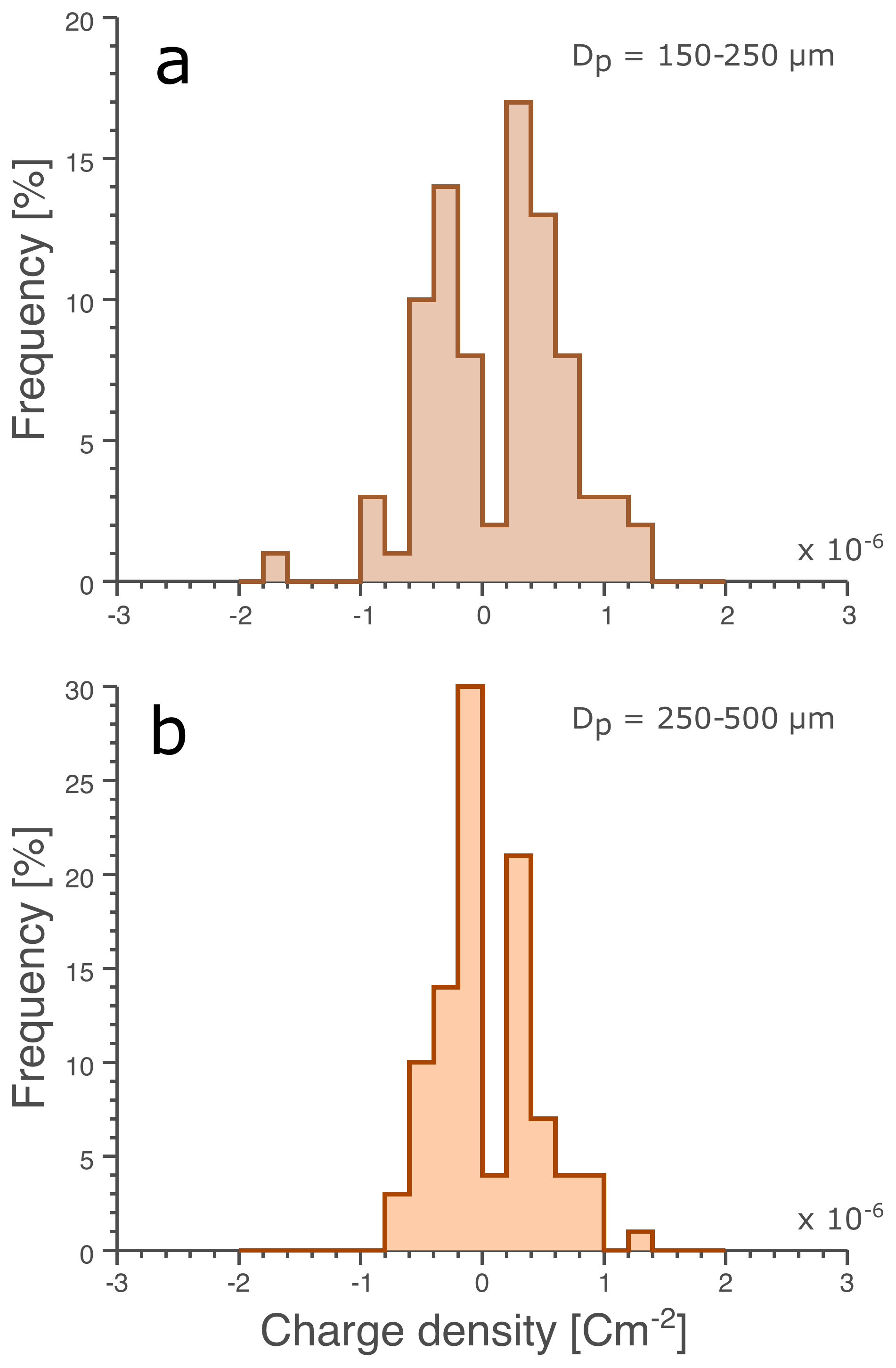}
	\caption{\label{fig9} Charge density distributions for particles of two nominal distributions a) 150-250 microns and b) 250-500 microns. Note that the troughs near zero are artifacts of a high noise floor.}
\end{figure}

The ability of a material to charge is often assessed by computing its surface charge density $\sigma$ (that is, the charge divided by a particle's surface area). The distributions of charge densities for the two granulometries are shown in \textbf{Figure \ref{fig9}}. In order to compare two size distributions, the charge densities here were calculated using the mean, spherical-equivalent particle diameter of each sample (200 $\mu$m for the 150-250 $\mu$m sample and 375 $\mu$m for the 250-500 $\mu$m sample). Under these assumptions, both size distributions gained maximum charge densities on the order of 10\textsuperscript{-6} Cm\textsuperscript{-2}. Numerous authors have noted that charging is enhanced in flows with broad particle size distributions (e.g. \cite{duff2008particle, forward_charge_2009}. Because our FPT setup restricted us to using narrow size distributions in the experiments where we measure charge density, the values we obtained for $\sigma$ may underestimate the range of charge densities in our full Martian dust simulant (which has particles smaller than 10 microns and as large as several millimeters). As discussed above, in an isolated flow, particles only exchange charge with each other. Thus, some grains gain net positive charge whereas others gain negative charge. This conservation of charge is reflected by the fact that the distributions are essentially Gaussian with zero means (See \textbf{Figure \ref{fig9}}). The troughs near the middle of the distributions (i.e. where the charge was very small) are artifacts of the high noise floor in our experiments. They do not represent true features of the distributions. 

\subsection{Charging rates and discharge frequency} \label{secResults:subsecRate}

\textcolor{black}{Previously, the charge on a particle in a granular medium as a function of time has been described by an exponential approach \citep{mendez_harper_effects_2016, mendez2020micro}:}

\begin{equation} \label{eq1}
    Q(t) = Q_{ss} + (Q_o - Q_{ss})/e^{t/\tau},
\end{equation}

\textcolor{black}{where $Q_{ss}$ is the steady-state charge, $Q_o$ is the initial charge, $\tau$ is the system time constant, and $t$ is time. The value of $\tau$ depends on particle collision rates, the efficiency of charge exchange during particle-particle contacts, as well as mechanisms that promote charge loss. Dividing \textbf{Equation \ref{eq1}} by the particle's surface area, allows us to investigate time-dependent charging in terms of surface charge density.}

\begin{equation} \label{eq2}
    \sigma(t) = \sigma_{ss} + (\sigma_o - \sigma_{ss})/e^{t/\tau},
\end{equation}

\textcolor{black}{At $t = 0$ (i.e. at the start of mobilization) the charge density is $\sigma_o$ and is typically taken to be 0 Cm\textsuperscript{-2}. After a long period ($>$ 5 time constants), the exponential term vanishes and the charge on the particle reaches its steady-state value $\sigma_{ss}$. To a first approximation $\sigma_{ss}$ equals the charge density on a particle just before breakdown, $\sigma_{\mbox{max}}$.}

\textcolor{black}{In the experiments with the free fountain, we observe that discharges begin to occur after 2 minutes of fluidization. The presence of discharges indicates that some particles in the system have gained $\sigma_{ss}$. Thus, the charging rate for these particles can be characterized by $\tau \approx$ 24 s. If we further assume that $\sigma_{ss} \sim $ 1e\textsuperscript{-6} Cm\textsuperscript{-2}--that is, the maximum charge densities measured in the fountain (\textbf{Figure \ref{fig9}})--, then \textbf{Equation \ref{eq2}} can be solved for all times. However, the fact that the number and magnitude of discharges in the fountain increases with time up until the electronics became saturated betrays the presence of longer time constants within the system. Such observation is consistent with the behavior of a similar spouted bed studied by \cite{mendez_harper_effects_2016}. That work reported time constants of 2 - 3 minutes for the triboelectric charging of ash volcanic ash under conditions comparable to those presented here. We postulate that these different timescales reflect the heterogeneous nature of the fluidized bed dynamics. The shorter time constant associated with the initiation of discharges reflect the fast charging of small clusters of particles entrained in the central jet of the fountain whereas the longer one is underpinned by the slower overturning dynamics throughout the entire particle bed.}

\textcolor{black}{As noted in \textbf{Section \ref{secResults:subsecdischarges}}, our experiments with an isolated flow generally produced fewer discharges per unit time (0.1-1 Hz) than previously published work and no visible glow discharge. For comparison, \cite{krauss_experimental_2003} report 0-15 discharges per second in experiments in which particles were agitated with a stirrer. Other work reports 0-4 discharges per second in experiments where mixtures of Martian dust simulants and glass microballoons were dropped from a hopper \citep{krauss2006modeling}. In these last experiments, discharges occurred within a second ($\sim$ 0.6 s) of the initiation of particle motion. Assuming the rate of charging in those experiments obeys the same exponential approach described in \textbf{Equation \ref{eq1}}, the system would be characterized by a much faster charging time constant of $\tau \approx$ 0.12 s.}

\textcolor{black}{The differences in discharge rates are likely underpinned by dynamical, environmental, and chemical dissimilarities between experiments \citep{lacks_contact_2011}. Firstly, because triboelectric charging requires physical contacts between grains, the charging rate (i.e. how fast the potential on a grain surface reaches the discharge threshold) depends on how often grains collide, the energy of particle-particle collisions, and contact time \citep{ireland_contact_2009, matsuyama2010maximum, liao2011effect, mendez_harper_effects_2016, jin2017role, grosshans2017direct, shinbrot2018multiple}. Such parameters, in turn, hinge on the number density of particles, a flow's granular temperature, and particle size \citep{goldhirsc2008, matsuyama2010maximum}. In general, triboelectric charging appears to benefit from greater collision rates, collision energies, and contact times \citep{mendez_harper_effects_2016} (although, recent work suggests that high contact frequencies may actually reduce the amount of charge under some conditions; e.g. \cite{shinbrot2018multiple}).}

\begin{figure}[!]
	\centering
	\includegraphics[width=4 in]{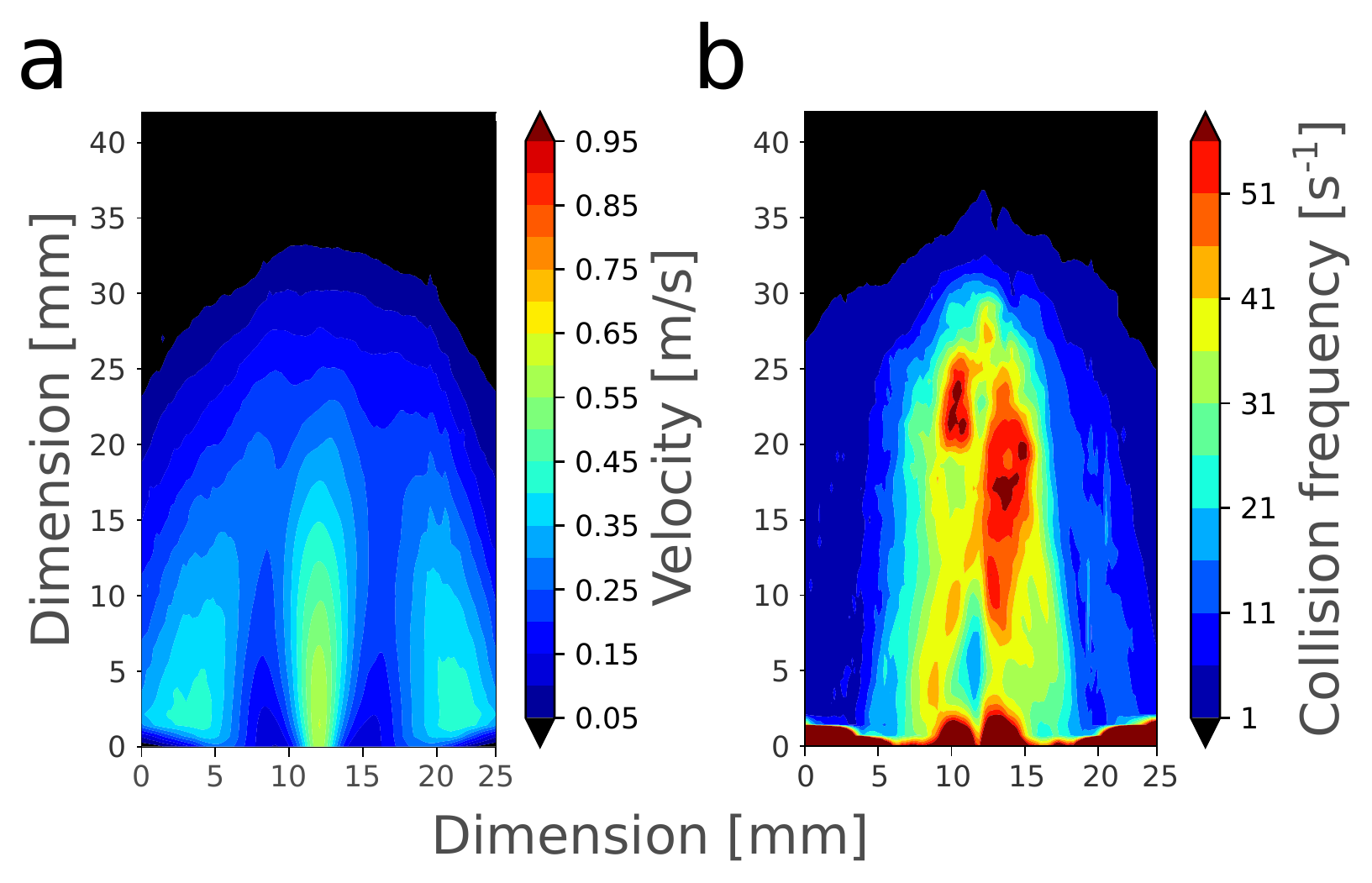}
	\caption{\label{fig10} Output of two fluid model. a) Velocity distribution within the fountain. Note the agreement with the velocities measure in the experimental fountain using particle image velocimetry. b) Collision frequency on per particle per second basis.}
\end{figure}

\textcolor{black}{A first order calculation in \cite{krauss2006modeling} indicates that those experiments involved rather dense particle flows, allowing for high collision frequencies and high charging rates. The authors estimate that each particle undergoes $\sim$ 8000 collisions a second-long experiment. Using a two-fluid numerical model (described within \textbf{\ref{secApendix}}), we estimate that the collision frequency in our spouted bed varies between $\sim$100 collisions per particle per second in the fountain's relatively dense central jet and 1 collision per particle per second in the dilute annulus (see \textbf{Figure \ref{fig10}}). On average, each particle in the fountain undergoes 20 collisions per second. The fact that the shortest charging time constant in our experiments ($\tau \sim$ 24 s) is two orders of magnitude larger than that in \cite{krauss2006modeling} ($\tau \sim$ 0.12 s) may be explained by collision frequencies two orders of magnitude higher in those experiments as compared to ours.}


\textcolor{black}{How fast a fluidized granular material charges also depends on the amount of charge exchanged per collision \citep{watanabe2006measurement, lee2018using}. As we have noted above, this amount may be higher for chemically-dissimilar mixtures with different work functions (e.g. \cite{eden1973electrical, krauss_experimental_2003, krauss2006modeling}) than for collections of chemically-identical particles (as is the case here and in \cite{wurm2019challenge}). Such hypothesis is supported by the fact that in experiments with the plastic insert the initiation of discharge occurs much sooner (at 30-40 seconds) than in those with an isolated fountain (2 min). Additionally, there is evidence that mobilizing particles at greater velocities produces a larger charge exchange per particle-particle collision (e.g. \cite{watanabe2006measurement}). Because we excited our particles such that larger particles would not be mobilized faster than the friction velocity at the surface, we expect charging to be subdued compared to experiments mobilizing simulants at higher velocities (for instance the free-stream wind velocity measured by the Viking Lander).}

\textcolor{black}{Lastly, the discharge frequency depends on how efficiently charge of opposite polarity can be separated to create the electric fields that lead to gas breakdown \citep{cimarelli_experimental_2014, yair2008charge}. In \cite{krauss2006modeling}, for instance, oppositely-charged, falling microballoons and natural silicates became separated by different settling velocities. In our small-scale spouted bed, however, the high-speed videography suggests that gravitational separation does not readily occur within the fountaining timescales. In other words, discharges in our fountain reflect electric fields generated by more stochastic arrangements of negative and positive particles. }

\textcolor{black}{In summary, we hypothesize that the low discharge rates in the isolated fountain and the absence of a visually-obvious plasma results from: 1) relatively low collision frequencies in the fountain; 2) the lack of chemically-dissimilar contacts; 3) the low-energy contacts between particles; and 4) fountaining dynamics that do not promote spatial separation between charged particles of different polarities.}

\subsection{Implications for Martian atmospheric electricity} 

Electrical discharges have not been incontrovertibly detected on Mars. Lacking field data, most work has focused on determining i) a viable electrification mechanism capable of charging Martian materials, ii) establishing criteria for the breakdown of the Martian atmosphere, and iii) gaining insight from terrestrial dust systems. Given the lack of precipitation on Mars, any electrical activity would likely be underpinned by electrification mechanisms other than those believed to operate within conventional thunderstorms (i.e those involving ice and graupel; \cite{keith1990lab, saunders1991effect}). Tribocharging has been proposed repeatedly as a viable electrification mechanism on Mars given its generally accepted roles in terrestrial dust flows \citep{eden1973electrical, mills1977dust, krauss_experimental_2003, merrison2004electrical, krauss2006modeling, kok_wind_blown_2009, farrell_is_2015, harrison2016applications}. It is interesting to note, however, that while numerous authors report large electric fields associated with terrestrial dust systems, only a handful describe electrical discharges--e.g. \cite{kamra1972electric}. This dearth in reporting may stem from the more stringent breakdown criteria associated with the near-surface terrestrial atmosphere.  

Yet, even if triboelectric processes can statically charge aeolian materials, some work has cast doubt as to whether frictional interactions are efficient enough to cause atmospheric breakdown on Mars. For instance, \cite{kok_wind_blown_2009} have suggested that locally generated plasma in dust storms may increase the conductivity of the gas, therefore "short-circuiting" the electric field set up by charged particles. This quenching could prevent electric fields from raising above the breakdown threshold, complicating the production of arc discharges. \cite{farrell_is_2015} has countered by showing that the current densities associated with pre-spark Townsend discharge (electron avalanche) are generally smaller than the triboelectric charging currents. Although the ``short circuiting" currents may become comparable to the charging currents when the electric fields in the dust storm become large, they appear to be incapable of shutting off the discharge completely. \textcolor{black}{Indeed, \cite{harrison2016applications} indicate that, while large-scale concentrations of charge are unlikely on Mars, the conductivity of the Martian atmosphere within dust clouds may be sufficiently low to accommodate localized spark and glow discharges.}  More recently, \cite{riousset2020scaling} coupled a Boltzmann solver with standard Global Reference Atmospheric Models to estimate the breakdown conditions at every altitude for realistic gas mixtures on Mars. These authors find that the low pressure Martian atmosphere does indeed favor discharges. 

Our experiments allow us to address some of these open questions. While electrification processes other than tribocharging could operate in global dust storms and dust devils, the presence of discharges in our fluidized bed experiment supports the hypothesis that triboelectrification alone may meet the criteria for the production of small spark discharges under Martian conditions. Furthermore, the charge densities we compute by directly measuring the charge on individual grains is generally consistent with the theoretical criteria for breakdown. At the surface, the maximum electric field $E_{\mbox{max}}$ that can be sustained by the tenuous Martian atmosphere  has been estimated to fall between 20-50 kV/m \citep{melnik1998electrostatic, kok_wind_blown_2009}. From the electric field, the maximum charge density can be computed to a first order as:

\begin{equation} \label{eq3}
    \sigma_{\mbox{max}} = \epsilon_{0} E_{\mbox{max}},
\end{equation}

where $\epsilon_0$ is the permitivitty, and  $E_{\mbox{max}}$ is the breakdown electric field. Thus, theoretically, $\sigma_{\mbox{max}}$  ranges between 2.8 and 5 $\times$ 10\textsuperscript{-7} Cm\textsuperscript{-2} (for comparison, the $\sigma_{\mbox{max}}$ on Earth is 27 $\times$ 10\textsuperscript{-5} Cm\textsuperscript{-2}). Note that the charge densities we measure in both granulometries meet this discharge criteria (see \textbf{Figure \ref{fig9}}). In fact, a fraction of the particles we characterized carried charge densities that exceed these theoretical values by up to a factor of 5. \textcolor{black}{We posit that the large values of charge density may result from an underestimate of the particle surface area. As noted previously, we used the spherical equivalent diameter to compute these areas. However, a highly irregular particle of a given size is likely to have a larger surface areas than its spherical counterpart. In other words, for a given amount of charge, an irregular particle would tend to have a lower surface charge density than a spherical grain. Another source of error in the estimate of the surface charge density may have to do with \textbf{Equation \ref{eq3}} itself.} \cite{hamamoto_experimental_1992} argues that charge densities that exceed the theoretical limit (\textbf{Equation \ref{eq3}}) are to be expected on on sub-millimeter particles because the region of high electric field in which electron avalanche occurs becomes thinner as the rate of curvature (as grains get smaller) increases. Indeed, those authors find that particles around 1 micron may sustain charge densities several orders of magnitudes larger than that predicted by equation \ref{eq1}. This may also explain why the charge density distribution for the 150-250 $\mu$m sample is broader than that for the 250-500 $\mu$m sample.

Both the presence of discharges in the active fountain and the elevated charge densities on individual grains suggest that triboelectric charging is indeed capable of meeting the breakdown criteria for near-surface Martian conditions. This conclusion is consistent with the Paschen Law limit model discussed in \cite{wurm2019challenge}. \cite{wurm2019challenge} hypothesized that minute discharges between grains occurred to limit the amount of charge collected by grains. Here, we have explicitly shown that such small-scale discharges do indeed occur in isolated flows of natural silicate materials under a simulated Martian environment moving at the frictional threshold velocity, as well as quantified the charge densities that the Mars simulants reach.

We stress, as other have done before (e.g. \cite{wurm2019challenge}), that the presence of small-scale discharges does not imply the existence of large-scale lightning on Mars. Lightning not only requires an efficient charging mechanism to electrify particles, but also processes by which negative and positive charge become separated over distances of hundreds to thousands of meters.  Perhaps the best terrestrial analog to putative Martian discharges are the small volcanic vent discharges that occur in the rarefied barrel shock region of a supersonic volcanic jet \citep{behnke_observations_2013, mendez2018inferring}. These small discharges (less than 10 m in length) are only present within the region of low pressure just upstream of a Mach disk. Because vent discharges involve low energies, they are invisible to global lightning detection systems, but can be detected from a few kilometers away using specialized equipment like a lightning mapping array. Even close to the source, vent discharges manifest not as discrete pulses in radio frequency, but rather as a continual RF ``hum'' (see \textbf{Figure 1} in \cite{thomas_electrical_2007}). If discharges in Martian dust storms are comparable to volcanic vent discharges, they would likely be undetectable from orbit. Positive detection would require dedicated instruments on the ground.

\section{Summary}

In this work, we have investigated the capacity of a Martian dust simulant to charge triboelectrically in an isolated granular flow. Unlike previous work, our experiments were designed to minimize particle contacts with foreign surfaces, as well as designed to measure individual spark discharges.

We have demonstrated that:

(i) Natural basaltic grains  in an isolated (i.e. with no wall contacts) spouted bed under near-surface Martian conditions and velocities that approximate the frictional threshold can produce spark discharges. 

(ii) The distribution of charge densities on the particles were measured, with maximum charge densities on the order of 10\textsuperscript{-6} Cm\textsuperscript{-2}. 

(iii) Grains coming into contact with a chemically-dissimilar laboratory container affects the polarity with which dust grains charge and influences discharge behavior.

Forty years of laboratory experiments and modelling have suggested that the small-scale discharges may occur within Mar's near-surface dust system. Nevertheless, most of those experiments involved interactions between Martian dust simulant grains and container walls. Such contacts could have produced electrification behaviors different from those on Mars. Here, we have directly accounted for these concerns and shown that spark discharges do indeed still occur within an isolated flow of basaltic grains. These discharges are small in scale and we suggest an analogy to volcanic vent discharges on Earth. Given their low energies, we will likely have to wait for on-ground electrical measurements to unequivocally ascertain whether discharges occur in granular flows on the actual Martian surface. If electrical discharges do exist on Mars, it will be interesting to quantify whether the sizes of discharges scale with the size of the dust devil or storm and whether or not they can catalyze chemical reactions. 

\section{Acknowledgements}

The authors would like to acknowledge the Blue Waters Graduate Fellowship for support of JMH and grant EAR 1645057 for JD.

 \appendix

 \section{Two fluid numerical model}
\label{secApendix}

\textcolor{black}{To estimate the collision rate in our spouted bed, we used a numerical model together with kinetic theory to estimate the rate of triboelectric charging \citep{gidaspow1994multiphase}. Specifically, we employed  the Multiphase Flow with Interphase eXchanges (MFiX) code developed by the National Energy Technology Laboratory (NETL). The MFiX code leverages a 2-dimensional, two-fluid approach, in which the solid and gas phase in the fountain are treated as interpenetrating media with separate conservation equations governing mass balance and momentum. The two phases are coupled through interaction terms in these equations. Because the solid phase is treated as a continuum, the forces on individual grains in the fountain are not computed. Instead, the model averages solid and fluid properties over each cell of a discretized domain.} 

\textcolor{black}{The geometry of the simulation domain was tailored to that of our physical setup (roughly 7 by 7 cm). We used a solid phase with a density of 2800 kg m\textsuperscript{-3} and a particle size of 75 $\mu$m. This particle size was selected because grains with diameters on the order of 10\textsuperscript{-5} m are the most abundant in our Martian dust simulant (see \textbf{Section \ref{secDust}}). We validate the model by comparing the time-averaged morphology and velocity distributions of simulated fountain to those of the experimental fountain. The distribution of velocity magnitudes in the simulated fountain, averaged over 10 seconds of simulation time, is rendered in \textbf{Figure \ref{fig10}a}. Note the agreement between the velocity distribution in the fountain calculated numerically (\textbf{Figure \ref{fig10}}a) and that extracted through particle image velocimetry (\textbf{Figure \ref{fig2}b}). Specifically, the maximum velocities near the gas inlet ($\sim$ 0.6 m/s), as well as the shape and size of the fountain, are generally reproduced by the model.}

\textcolor{black}{The main piece of information extracted from the model is the fountain's granular temperature $\theta$. This last quantity is defined as $\theta = 1/2 < c_i^2>$, where $c_i$ is the fluctuating component of the particle velocity in either the horizontal or vertical direction, $i = x,y$ \citep{cody1996particle}.  Subsequently, the average number of collisions a single particle experiences per second can be computed as \citep{goldhirsc2008, aguilar2011collisions}:}

\begin{equation}
    f = \frac{12 \epsilon_s g_0}{\sqrt{\pi} D_p} \sqrt{\theta}.
\end{equation}

\textcolor{black}{Above, $D_p$ is the particle size. The radial distribution function $g_0$ depends on $\epsilon_s$ and the close-packed solid fraction $\epsilon_{s0} \sim 0.64$ \citep{aguilar2011collisions}:}

\begin{equation}
   g_0 = \left[1 - \left(\frac{\epsilon_s}{\epsilon_{s0}}\right)^{1/3}\right]^{-1}.
\end{equation}

\textcolor{black}{The average collisional frequency $f$ for the particles in the fountain across 10 seconds of simulation time is shown in \textbf{Figure \ref{fig10}b}. In the core of the fountain, $f$ takes maximum values of $\sim$100 s\textsuperscript{-1} particle\textsuperscript{-1}, whereas each particle in the annular region undergo fewer than 10 collisions per second. The average number of contacts a particle experiences per second is $\sim 20$. These values are consistent with previous work on the collisional dynamics in spouted beds with similar fluidization and granular parameters (e.g \cite{tamrakar2018advanced}).}




\section{References}
\bibliography{references}

\end{document}